\documentclass[onecolumn, doublespace, 12pt fonts]{aastex6}

\begin{document}

\title{Neutron star equation of state from the quark level in the light of GW170817}
\author{Zhen-Yu Zhu\altaffilmark{1}, En-Ping Zhou\altaffilmark{2}, Ang Li\altaffilmark{1}}

\altaffiltext{1}{Department of Astronomy, Xiamen University, Xiamen, Fujian 361005, China; liang@xmu.edu.cn}
\altaffiltext{2}{State Key Laboratory of Nuclear Science and Technology and School of Physics, Peking University, Beijing 100871, China}

\begin{abstract}
Matter state inside neutron stars is an exciting problem in astrophysics, nuclear physics and particle physics. The equation of state (EOS) of neutron stars plays a crucial role in the present multimessenger astronomy, especially after the event of GW170817. We propose a new neutron star EOS ``QMF18'' from the quark level, which describes well robust observational constraints from free-space nucleon, nuclear matter saturation, heavy pulsar measurements and the tidal deformability of the very recent GW170817 observation. For this purpose, we employ the quark-mean-field (QMF) model, allowing one to tune the density dependence of the symmetry energy and study effectively its correlations with the Love number and the tidal deformability. We provide tabulated data for the new EOS and compare it with other recent EOSs from various many-body frameworks.
\end{abstract}


\section{Introduction}
%------------------------------------

Neutron stars (NSs) are by far one of the most interesting observational objects, since many mysteries remain on them due to their complexity. Multimessenger observations with advanced telescopes such as Advanced LIGO and VIRGO~\citep[e.g.,][]{2017PhRvL.119p1101A} , FAST~\citep[e.g.,][]{2016RaSc...51.1060L}, SKA~\citep[e.g.,][]{2015aska.confE..43W},  NICER~\citep[e.g.,][]{2016ApJ...832...92O}, HXMT~\citep[e.g.,][]{2018SCPMA..61c1011L}, eXTP~\citep[e.g.,][]{extp}, AXTAR~\citep[e.g.,][]{2010SPIE.7732E..48R}, will hopefully provide precise measurements of their mass and/or radius, thus improving our current knowledge of such stellar objects and their equation of states (EOSs), especially for the high-density inner crust with densities above nuclear saturation density $\rho_0 \sim 0.16$ fm$^{-3}$. Dense matter EOS is also closely related to the scientific goals of all advanced radioactive beam facilities being built around the world ~\citep[e.g.,][]{2002Sci...298.1592D,2009PhRvL.102l2701T}.

Nowadays, the EOS of symmetric nuclear matter (SNM)~\citep{2002Sci...298.1592D} ($\beta\equiv\frac{\rho_n-\rho_p}{\rho_n+\rho_p}=0$) is relatively well-constrained, with $\rho_n$, $\rho_p$, the neutron, proton density, respectively. Matter with nonzero isospin asymmetry remains unknown, largely due to the uncertainty in the symmetry energy: $E_{\rm sym}(\rho)\approx[E(\rho, \beta)-E(\rho, 0)]/\beta^2$, with $E(\rho, \beta)$ the energy per nucleon of nuclear matter at isospin asymmetry $\beta$ and density $\rho$. Conflicts remains for the symmetry energy (especially its slope $L(\rho)=dE_{\rm sym}(\rho)/d\rho$) despite significant progress in constraining the symmetry energy around and below nuclear matter saturation density~\citep[e.g.,][]{2009PhRvL.102l2701T,2014NuPhA.922....1D,2015PhRvC..92c1301Z}. At saturation density $\rho = \rho_0$, $L$ may has a lower limit $\sim 20$ MeV~\citep{2009PhRvL.102l2502C} and an upper limit $>170$ MeV~\citep{2013PhRvC..88d4912C}. It characterizes the density dependence of the
symmetry energy and largely dominates the ambiguity and stiffness of EOS for dense nuclear matter and NS matter at densities approached in NS cores, in the case of no strangeness phase transition~\citep[e.g.,][]
{2011PhRvC..83b5804B,2014PhRvC..89b5802H,2006PhRvC..74e5801L,2007ChPhy..16.1934L,2010PhRvC..81b5806L,2015PhRvC..91c5803L,2016PhRvC..94d5803Z}. Therefore it is a crucial parameter for NS EOS and related studies. 

Recently from the observation of GW170817, the LIGO+Virgo Collaborations placed a clean upper limit on the tidal deformability of the compact object, $\Lambda=(2/3)k_2/(GM/c^2R)^5$ with $k_2$ is the second Love number. Since the star radius is rather sensitive to the symmetry energy (essentially its slope $L$) with the maximum mass only slightly modified~\citep[e.g.,][]{2004Sci...304..536L,2001ApJ...550..426L,2006PhLB..642..436L}, possibly this $\Lambda$ measurement might put independent constraint on $L$, as has been previously discussed in \citet{2013PhRvC..87a5806F,2017arXiv171106615F,2018arXiv180106855Z}. $\Lambda$ describes the amount of induced mass quadrupole moment when reacting to a certain external tidal field \citep{1992PhRvD..45.1017D,2009PhRvD..80h4035D}. If a low spin prior is assumed for both stars in the binary, which is reasonable considering the magnetic braking during the binary evolution, the tidal deformability for a $1.4\,M_\odot$ star (denoted as $\Lambda$(1.4) in below) was concluded to be smaller than 800 (a more loosely constrained upper limit of 1400 is found for the high-spin prior case)~\citep{2017PhRvL.119p1101A}. Based on the GW170817 observation,
several recent studies have reported their constrains on NS EOS~\citep[e.g.,][]{2018arXiv180200571A,2017arXiv171106244A,2017arXiv171102644A,2017ApJ...850L..34B,2018ApJ...852L..32D,2017arXiv171106615F,2018arXiv180104620K,2017arXiv171105565M,2017ApJ...850L..19M,2018ApJ...857...12N,2018PhRvD..97h4038P,2018ApJ...852L..29R,2018PhRvD..97b1501R,2017PhRvD..96l3012S,2018arXiv180106855Z} and QS EOS~\citep{2017arXiv171104312Z}. 

The objective of the present study is to make use of the new GW170817 constraint, combined with available terrestrial nuclear structure/reaction experiments~\citep[e.g.,][]{2002Sci...298.1592D,2009PhRvL.102l2701T,2013PhLB..727..276L,2014NuPhA.922....1D,2015PhRvC..92c1301Z} and astrophysical observations~\citep{2013Sci...340..448A,2010Natur.467.1081D,2016ApJ...832..167F} for the determination of nuclear saturation properties, and to construct a new NS EOS from the quark level. The employed EOS model enables us to fine tune the $L$ value and study consistently the $\Lambda$ vs $L$ dependence, with well-reproduced robust observables from laboratory nucleons, nuclear saturation, heavy-ion collisions (HIC) and heavy pulsars.

The paper is organized as follows. In Section 2, we describe the theoretical framework to describe consistently a nucleon and many-body nucleonic system from a quark potential, including the necessary fitting of the quark potential parameters and the meson coupling parameters from vacuum nucleon and empirical nuclear saturation properties, respectively. In Section 3, the NS EOSs and the corresponding mass-radius relations as well as the tidal deformabilities are discussed, to be compared with the mass measurements of heavy pulsars~\citep{2013Sci...340..448A,2010Natur.467.1081D,2016ApJ...832..167F} and the tidal deformability of GW170817 (either $\Lambda(1.4)$ or $\tilde{\Lambda}$)~\citep{2017PhRvL.119p1101A}; Tabulated EOS of the new ``QMF18'' model is also provided and comparisons are made with other recent EOSs from various many-body theories. Summary and future perspective finally presented in Section 4.

\section{Model}
%------------------------------------
To carry out a study of nuclear many-body system from the quark level, one first constructs a nucleon from confined quarks, by a finite confining region (characterized by a constant energy per unit volume, the bag constant $B$)~\citep{1974PhRvD...9.3471C} or by constituent quarks with a harmonic oscillator confining potential~\citep{1986PhRvD..33.1925B,1989JPhG...15..297F}. Then nucleons interact with point-like mesons. Since the meson fields modify the internal quark motion, the mesons couple not to point-like nucleons but self-consistently to confined quarks. The effect of the nucleon velocity as well as the effect of antisymmetrisation are usually neglected, and the calculation is done in the mean-field approximation. The first model is often called the quark-meson coupling (QMC)~\citep[e.g.,][]{2013PhRvC..88a5206B,1988PhLB..200..235G,2015PhRvC..92d5203M,2016PhRvC..94c5805M,2007PrPNP..58....1S}, and the later is called the quark mean-field (QMF) model~\citep[e.g.,][]{2014PTEP.2014a3D02H, 2014PhRvC..89b5802H,2000PhRvC..61d5205S,2002NuPhA.707..469S,1998PhRvC..58.3749T,2016PhRvC..94d4308X,2018arXiv180207441Z}. The QMC and QMF models may be viewed as variation of the relativistic mean-field (RMF) model~\citep[e.g.,][]{1977NuPhA.292..413B,2001PhRvL..86.5647H,1996NuPhA.606..508M,rmf86,1974AnPhy..83..491W} which is from the hadron level. The RMF model, including its extension of the isoscalar Fock terms~\citep[e.g.,][]{2018arXiv180107084J,2012PhRvC..85b5806L,2008PhRvC..78f5805S,2016PhRvC..94d5803Z}, has been widely-used for NS studies.

In this work, following the methodology of the QMC and QMF models, we start with a flavor independent potential $U(r)$ confining the constituent quarks inside a nucleon. Details can be found in~\citet{2013PhRvC..88a5206B, 2015PhRvC..92d5203M, 2016PhRvC..94c5805M,2016PhRvC..94d4308X,2018arXiv180207441Z}. Here for completeness, we only write necessary formulas. The confining potential is written as~\citep{1986PhRvD..33.1925B}:
\begin{eqnarray}
U(r)=\frac{1}{2}(1+\gamma^0)(ar^2+V_0),
\end{eqnarray}
with the parameters $a$ and $V_0$ to be determined from vacuum nucleon properties. 
The Dirac equation of the confined quarks is written as
\begin{eqnarray}
[\gamma^{0}(\epsilon_{q}-g_{\omega q}\omega-\tau_{3q}g_{\rho q}\rho)-\vec{\gamma}\cdot\vec{p} -(m_{q}-g_{\sigma q}\sigma)-U(r)]\psi_{q}(\vec{r})=0,
\end{eqnarray}
Hereafter $\psi_{q}(\vec{r})$ is the quark field, $\sigma$, $\omega$, and $\rho$ are the classical meson fields. $g_{\sigma q}$, $g_{\omega q}$, and $g_{\rho q}$ are the coupling constants of $\sigma, ~\omega$, and $\rho$ mesons with quarks, respectively. $\tau_{3q}$ is the third component of isospin matrix. This equation can be solved exactly and its ground state solution for energy is
\begin{eqnarray}
(\mathop{\epsilon'_q-m'_q})\sqrt{\frac{\lambda_q}{a}}=3,
\end{eqnarray}
where $\lambda_q=\epsilon_q^\ast+m_q^\ast,\ \mathop{\epsilon'_q}=\epsilon_q^\ast-V_0/2,\ \mathop{m'_q}=m_q^\ast+V_0/2$. The effective single quark energy is given by $\epsilon_q^*=\epsilon_{q}-g_{q\omega}\omega-\tau_{3q}g_{q\rho}\rho$ and the effective quark mass by $m_q^\ast = m_q-g_{\sigma q}\sigma$ with the quark mass $m_q$ = 300 MeV.

The zeroth-order energy of the nucleon core $E_N^0=\sum_q\epsilon_q^\ast$ can be obtained by solving Eq.~(3). The contribution of center-of-mass correction $\epsilon_{c.m.}$, pionic correction $\delta M_N^\pi$ and gluonic correction $(\Delta E_N)_g$ are also taken into account as done in~\citet{2013PhRvC..88a5206B,1986PhRvD..33.1925B,2015PhRvC..92d5203M,2016PhRvC..94c5805M,2016PhRvC..94d4308X,2018arXiv180207441Z}. With these corrections on energy, we can then determine the mass of nucleon in medium:
\begin{eqnarray}
M^\ast_N=E^{0}_N-\epsilon_{c.m.}+\delta M_N^\pi+(\Delta E_N)_g.
\end{eqnarray}
The nucleon radius is written as
\begin{eqnarray}
\langle r_N^2\rangle = \frac{\mathop{11\epsilon'_q + m'_q}}{\mathop{(3\epsilon'_q + m'_q)(\epsilon'^2_q-m'^2_q)}}.
\end{eqnarray}
From reproducing the nucleon mass and radius $(M_N, r_N)$ in free space, we determine the potential parameters ($a$ and $V_0$) in Eq.~(1). $V_0=-62.257187$ MeV and $a=0.534296$ fm$^{-3}$ are obtained by fitting $M_N = 939$ MeV and $r_N = 0.87$ fm.

We then move from a single nucleon to nucleonic many-body system for the study of infinite nuclear matter and NSs. Nuclear matter is described by point-like nucleons and mesons interacting through exchange of $\sigma,~\omega,~\rho$ mesons. The Lagrangian is written as (see also for example in \citet{2008PhR...464..113L}):
\begin{eqnarray}
\mathcal{L}& = & \overline{\psi}\left(i\gamma_\mu \partial^\mu - M_N^\ast - g_{\omega N}\omega\gamma^0 - g_{\rho N}\rho\tau_{3}\gamma^0\right)\psi  -\frac{1}{2}(\nabla\sigma)^2 - \frac{1}{2}m_\sigma^2 \sigma^2 - \frac{1}{3} g_2\sigma^3 - \frac{1}{4}g_3\sigma^4 \nonumber \\ 
& & + \frac{1}{2}(\nabla\rho)^2 + \frac{1}{2}m_\rho^2\rho^2 + \frac{1}{2}(\nabla\omega)^2 + \frac{1}{2}m_\omega^2\omega^2 + \frac{1}{2}g_{\rho N}^2\rho^2 \Lambda_v g_{\omega N}^2\omega^2, 
\end{eqnarray}
where $g_{\omega N}$ and $g_{\rho N}$ are the nucleon coupling constants for $\omega$ and $\rho$ mesons. From the quark counting rule, we obtain $g_{\omega N}=3g_{\omega q}$ and $g_{\rho N}=g_{\rho q}$. The calculation of confined quarks gives the relation of effective nucleon mass $M_N^*$ as a function of $\sigma$ field, which defines the $\sigma$ coupling with nucleons (depending on the parameter $g_{\sigma q}$). $m_{\sigma} = 510~\rm{MeV}$,~$m_{\omega}=783~\rm{MeV}$, and $m_{\rho}=770~\rm{MeV}$ are the meson masses. 

\begin{table}\label{tab1}
\begin{center}
\caption{Saturation properties used in this study for the fitting of new sets of meson coupling parameters: The saturation density $\rho_0$ (in fm$^{-3}$) and the corresponding values at saturation point for the binding energy $E/A$ (in MeV), the incompressibility $K$ (in MeV), the symmetry energy $E_{\rm sym}$ (in MeV), the symmetry energy slope $L$ (in MeV) and the ratio between the effective mass and free nucleon mass $M_N^\ast/M_N$.}
\begin{tabular}{c|ccccc}\hline
$\rho_0$ & $E/A$  & $K$ & $E_{\rm sym}$& $L$ & $M_N^\ast/M_N$ \\ 
$[{\rm fm}^{-3}]$ & [MeV] & [MeV] & [MeV] & [MeV] & / \\ \hline
0.16 & -16 & 240 & 31 &20/40/60/80 & 0.77 \\ \hline
\end{tabular}
\end{center}
\end{table}
\begin{table*}\label{tab2}
\begin{center}
\caption{Newly fitted meson coupling parameters by using Table 1 as input.}
\begin{tabular}{cccccccc}\hline
$L$ [MeV]& $g_{\sigma q}$ & $g_{\omega q}$ & $g_{\rho q}$ & $g_2$ [fm$^{-1}$] & $g_3$ & $\Lambda_v$ \\ \hline
20 & 3.8620366 & 2.9174838 & 6.9588083 & 14.6179599 & -66.3442468 & 1.1080665 \\
40 & 3.8620366 & 2.9174838 & 5.4129448 & 14.6179599 & -66.3442468 & 0.7693664 \\
60 & 3.8620366 & 2.9174838 & 4.5830609 & 14.6179599 & -66.3442468 & 0.4306662 \\
80 & 3.8620366 & 2.9174838 & 4.0459574 & 14.6179599 & -66.3442468 & 0.0919661 \\ \hline
\end{tabular}
\end{center}
\end{table*}

The equations of motion for mesons can be obtained by variation of the Lagrangian:
\begin{eqnarray}
&&(i\gamma^{\mu}\partial_\mu-M_{N}^{\ast}-g_\omega\omega\gamma^0-g_\rho\rho\tau_3\gamma^0)\psi=0,\\
&&m_{\sigma}^2\sigma+g_2\sigma^2+g_3\sigma^3=-\frac{\partial M_N^*}{\partial\sigma}\langle\bar{\psi}\psi\rangle,\\
&&m_{\omega}^2\omega+\Lambda_vg_{\omega N}^2g_{\rho N}^2\omega \rho^2=g_{\omega N}\langle\bar{\psi}\gamma^0\psi\rangle,\\
&&m_{\rho}^2\rho+\Lambda_vg_{\rho N}^2g_{\omega N}^2\rho \omega^2=g_{\rho N}\langle\bar{\psi}\tau_3\gamma^0\psi\rangle.
\end{eqnarray}
From these Lagrangian and equations of motion of nucleon and mesons, the energy density and pressure can be generated by the energy-momentum tensor:
\begin{eqnarray}
\mathcal{E}&=&\frac{1}{\pi^2}\sum_{i=n,p}\int^{k^i_F}_0\sqrt{k^2+M_N^{\ast2}}k^2dk 
+\frac{1}{2}m^2_\sigma\sigma^2+\frac{1}{3}g_2\sigma^3+\frac{1}{4}g_3\sigma^4 \nonumber \\ 
&&+\frac{1}{2}m^2_\omega\omega^2+\frac{1}{2}m^2_\rho\rho^2+ \frac{3}{2}\Lambda_vg_{\rho N}^2g_{\omega N}^2\rho^2\omega^2,
\end{eqnarray}
\begin{eqnarray}
P & = & \frac{1}{3\pi^2}\sum_{i=n,p}\int_0^{k_F^i}\frac{k^4}{\sqrt{k^2+M_N^{\ast2}}}dk - \frac{1}{2}m_\sigma^2\sigma^2-\frac{1}{3}g_2\sigma^3-\frac{1}{4}g_3\sigma^4 \nonumber \\
& & + \frac{1}{2}m_\omega^2\omega^2 + \frac{1}{2}m_\rho^2\rho^2 + \frac{1}{2}\Lambda_vg_{\rho N}^2g_{\omega N}^2\rho^2\omega^2.
\end{eqnarray}
$k_F^p~(k_F^n$) is the Fermi momentum for proton~(neutron).

There are six parameters ($g_{\sigma q}, g_{\omega q}, g_{\rho q}, g_2, g_3, \Lambda_v$) in the Lagrangian of Eq.~(6) and they will be determined by fitting the saturation density $\rho_0$ and the corresponding values at saturation point for the binding energy $E/A$, the incompressibility $K$, the symmetry energy $E_{\rm sym}$, the symmetry energy slope $L$ and the effective mass $M_N^\ast$. Those employed values are collected in Table 1. We use the intermediate value of incompressibility $K \approx 240\pm20$ MeV from \citet{2010JPhG...37f4038P,2006EPJA...30...23S}. We also employ the most preferred values for ($E_{\rm sym},L$) newly suggested by ~\citet{2013PhLB..727..276L}, namely $E_{\rm sym} = 31.6\pm2.66$ MeV, $L \approx 58.9\pm16$ MeV. Since the $L$ value can be as low as $\sim$ 20 MeV~\citep{2009PhRvL.102l2502C}, we choose four values of $L$ (20, 40, 60, 80 MeV) as input of the parameter fitting according to our model capability, for studying its effect on the tidal deformatility of binary NS system~\citep{2017PhRvL.119p1101A}. The modal parameters obtained are collected in Table 2. Note that in our previous work~\citep{2018arXiv180207441Z}, with similar proper saturation properties ($\rho_0, E/A, K, E_{\rm sym}, L, M_N^\ast$) from terrestrial dense-matter measurements, the high-density EOS failed to reproduce the two-solar-mass-constraint from massive pulsars (only around 1.6 times the solar mass). For the present purpose of introducing new EOS for astrophysical studies, we refit the parameters by omitting the nonlinear terms of $\omega$ meson field and successfully obtain a maximum mass fulfilling the two-solar-mass constraint for the first time within QMF.

\section{EOS, mass-radius relation and tidal deformability} 

\begin{figure*}
\epsscale{1.0}
\includegraphics[width=20pc]{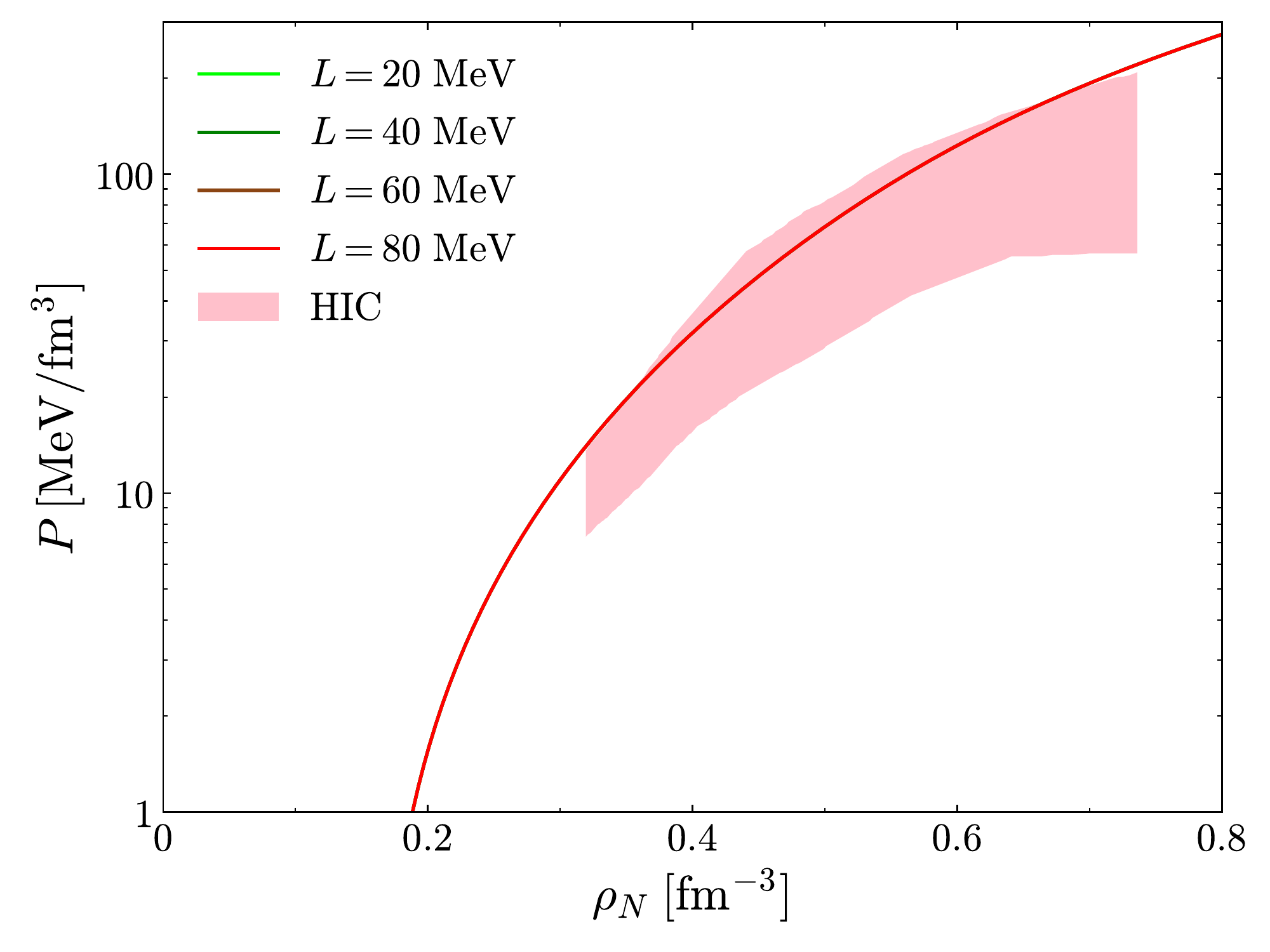}
\includegraphics[width=20pc]{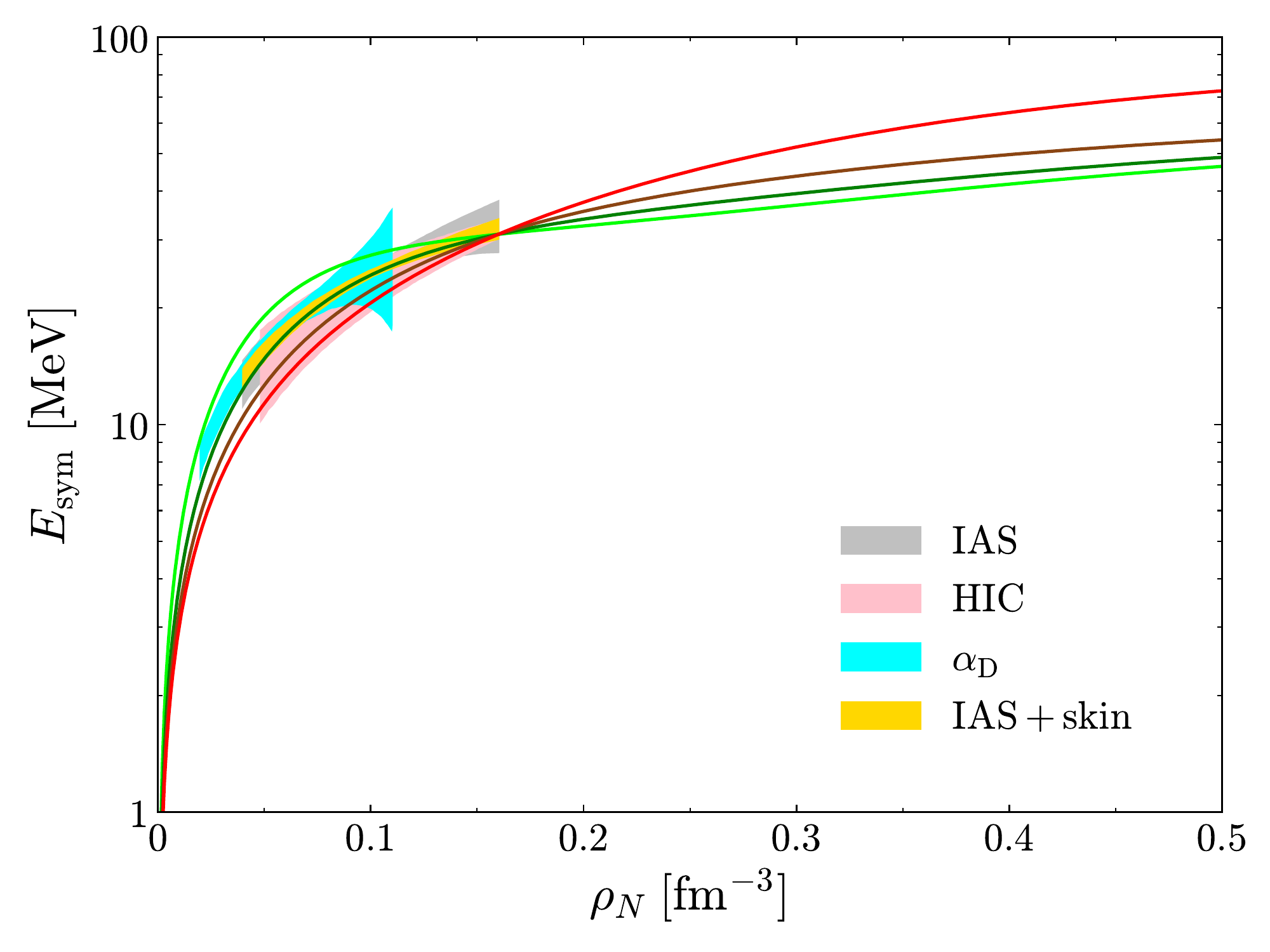}
\caption{(Left panel) Pressure as a function of nucleon number density for SNM, together with the constraint from collective flow in HIC~\citep{2002Sci...298.1592D} (Note here that the results from different parameter sets with different value of symmetry energy slope $L$ are the same in SNM); (Right panel) Symmetry energy as a function of nucleon number density with four different values of symmetry energy slope $L$. Colorful shadow regions represent the constraints from isobaric analog states (IAS) and from transport in HIC~\citep{2009PhRvL.102l2701T}, electric dipole polarizability in $^{\rm 208}$Pb ($\alpha_D$)~\citep{2015PhRvC..92c1301Z}, IAS and neutron skins (IAS+skin)~\citep{2014NuPhA.922....1D}, respectively.}\label{fig1}
\end{figure*}

After the meson coupling parameters are established, the pressure and symmetry energy as functions of density for nuclear matter can be calculated. The results are shown in Figure~1, together with experimental regions~\citep{2002Sci...298.1592D,2014NuPhA.922....1D,2009PhRvL.102l2701T,2015PhRvC..92c1301Z}. In the left panel, one can see that the SNM EOS is compatible with the flow constraint~\citep{2002Sci...298.1592D}. In the right panel for different $L$, the behaviour of symmetry energy vs density are all consistent with various nuclear experiments. Among them, the $L = 40$ MeV case lies comfortably inside all experimental boundaries.
 
\begin{figure*}
\epsscale{1.0}
\plotone{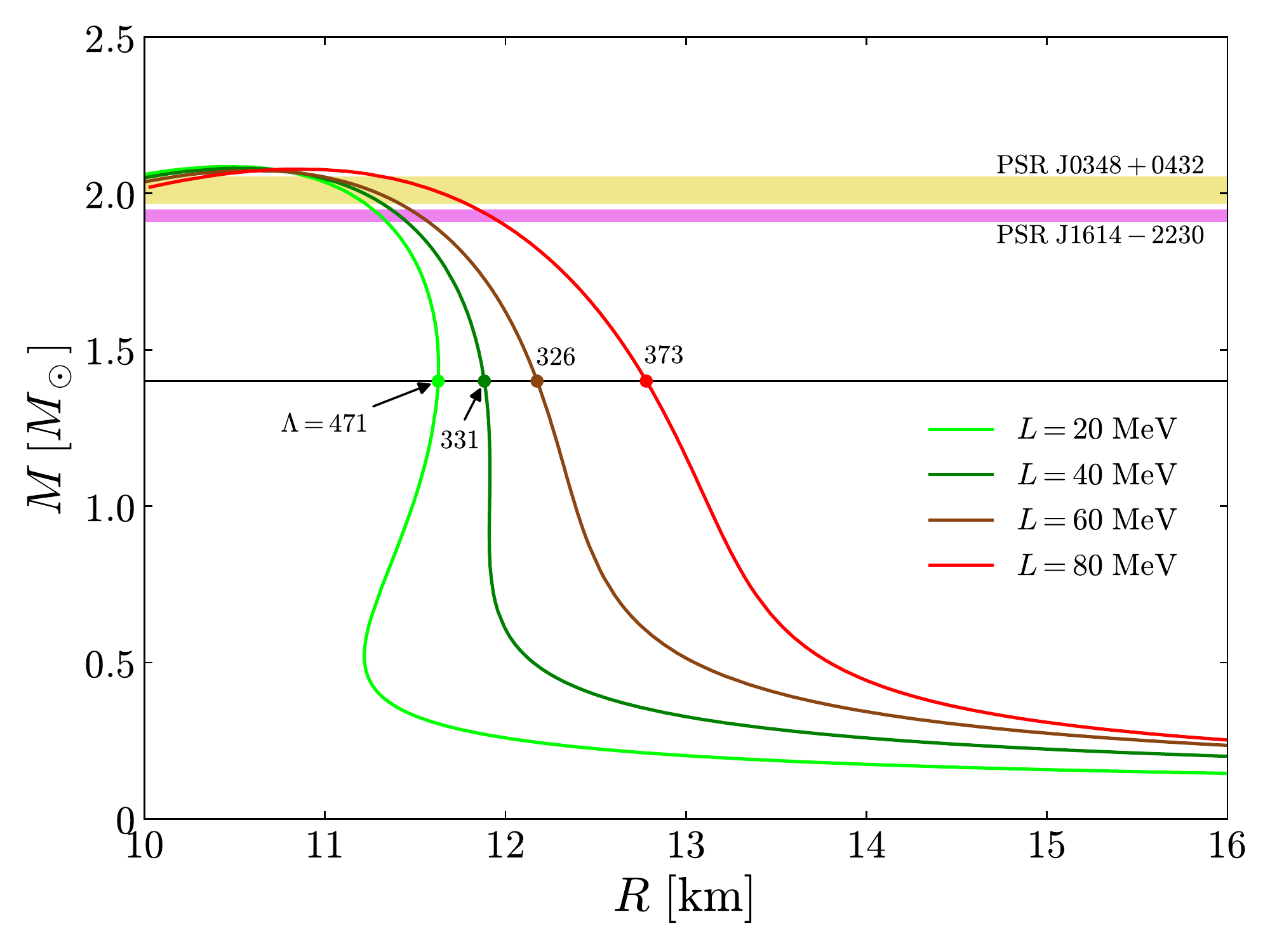}
\caption{Mass-radius curves for four EOSs with different value of $L$ ($20$, $40$, $60$, $80$MeV), together with the mass measurements for two recent massive stars: PSR J1614-2230~\citep{2010Natur.467.1081D,2016ApJ...832..167F} and PSR J0348+0432~\citep{2013Sci...340..448A}. The horizontal black line indicates $M = 1.4 \,M_\odot$. Numbers mark the $\Lambda$ values for $1.4\,M_\odot$ stars corresponding to colourful dots.}\label{fig2}
\end{figure*}

We can move forward to calculate the EOS of NS matter, $P(\mathcal{E})$, by introducing $\beta$-equilibrium and charge neutrality condition between nucleons and leptons:
\begin{eqnarray}
\mu_n = \mu_e + \mu_p,\ \ \ \rho_e + \rho_\mu = \mu_p,
\end{eqnarray}
where ($\mu_n$, $\mu_e$, $\mu_p$)/($\rho_n$, $\rho_e$, $\rho_p$) are the chemical potential/number density of neutron, electron and proton, respectively. To describe the structure of the crust, we employ the quantal calculations of ~\citet{nv} for the medium-density regime
($0.001\;\mathrm{fm}^{-3}<\rho_N<0.08\;\mathrm{fm}^{-3}$),
and follow the formalism developed in \citet{bps} for the outer crust ($\rho_N<0.001\;\mathrm{fm}^{-3}$). The tidal Love numbers $k_2$ is obtained from the ratio of the induced quadrupole moment $Q_{ij}$ to the applied tidal field $E_{ij}$~\citep{2009PhRvD..80h4035D,1992PhRvD..45.1017D,2008ApJ...677.1216H}:
$Q_{ij}=-k_2\frac{2R^5}{3G}E_{ij}$,
where $R$ is the NS radius. The dimensionless tidal deformability $\Lambda$ is related to the compactness $M/R$ and the Love number $k_2$ through $\Lambda = \frac{2}{3}k_2(M/R)^{-5}$. 

%fig3
\begin{figure}
\includegraphics[width=20pc]{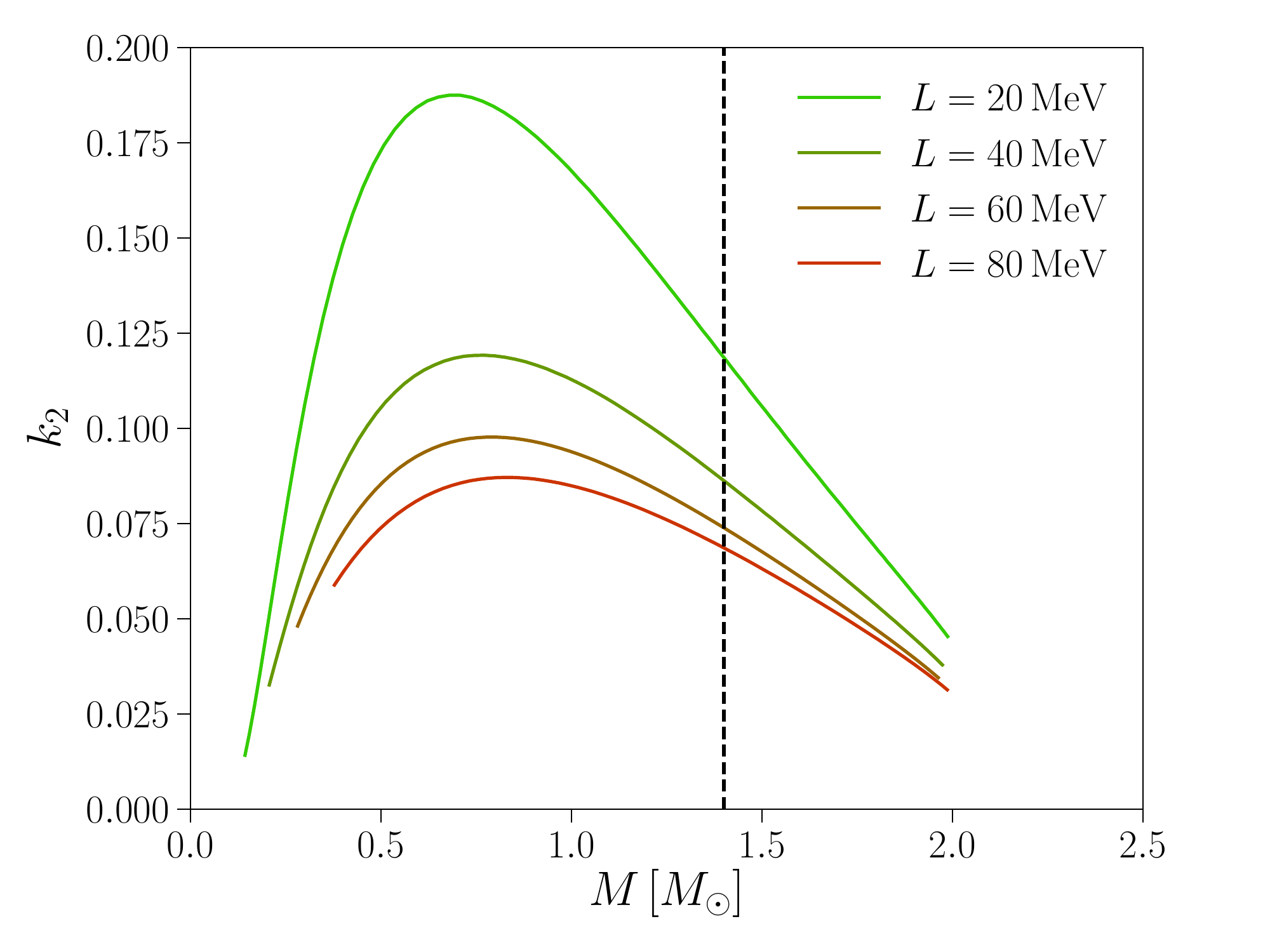}
\includegraphics[width=20pc]{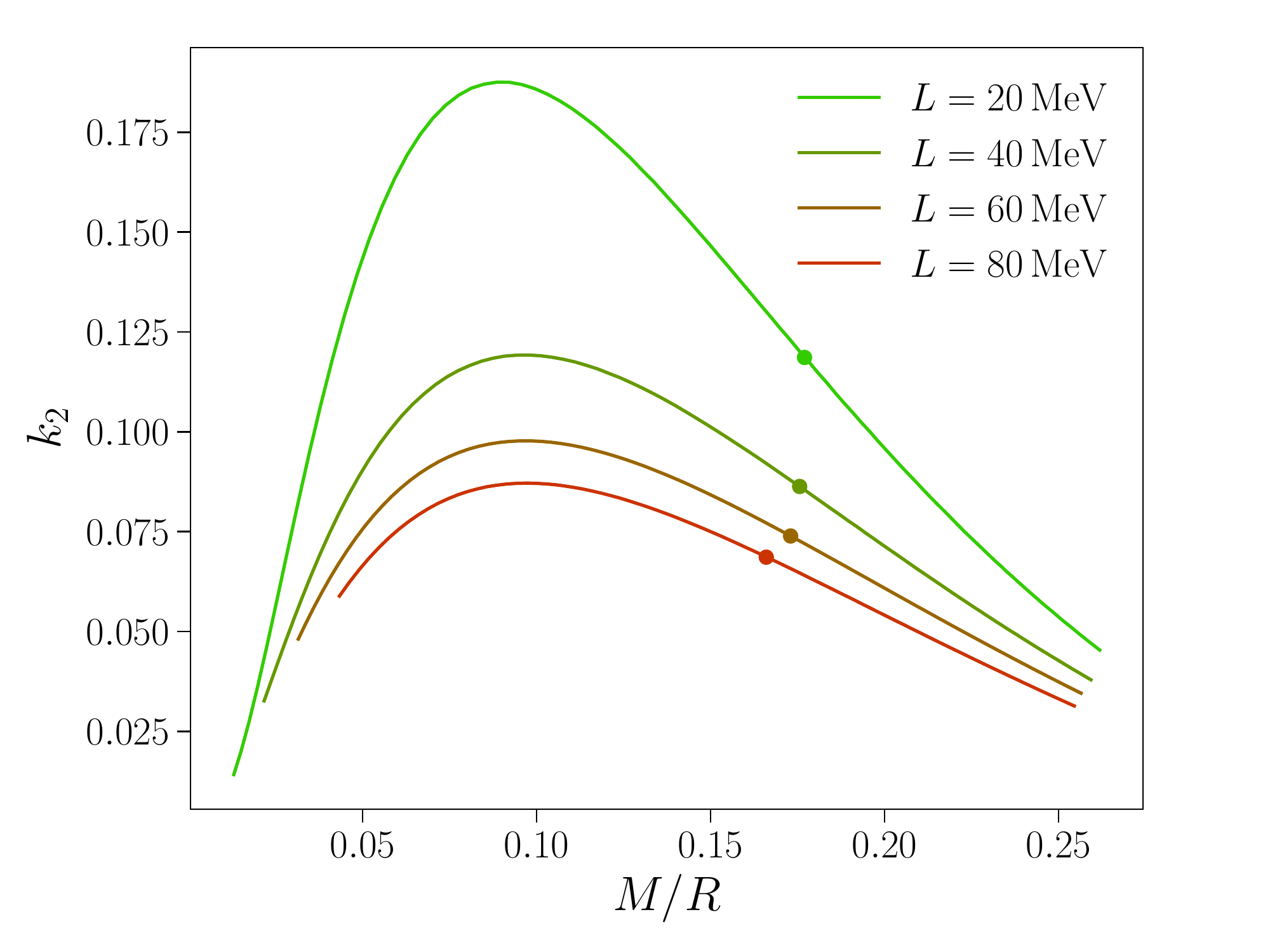}
\caption{Love numbers as a function of the mass (left panel) and the compactness (right panel), for four EOSs with different value of $L$ ($20$, $40$, $60$, $80$MeV). The vertical line and colorful dots indicate $M = 1.4\,M_\odot$.}
\end{figure}
%fig4
\begin{figure}
\includegraphics[width=20pc]{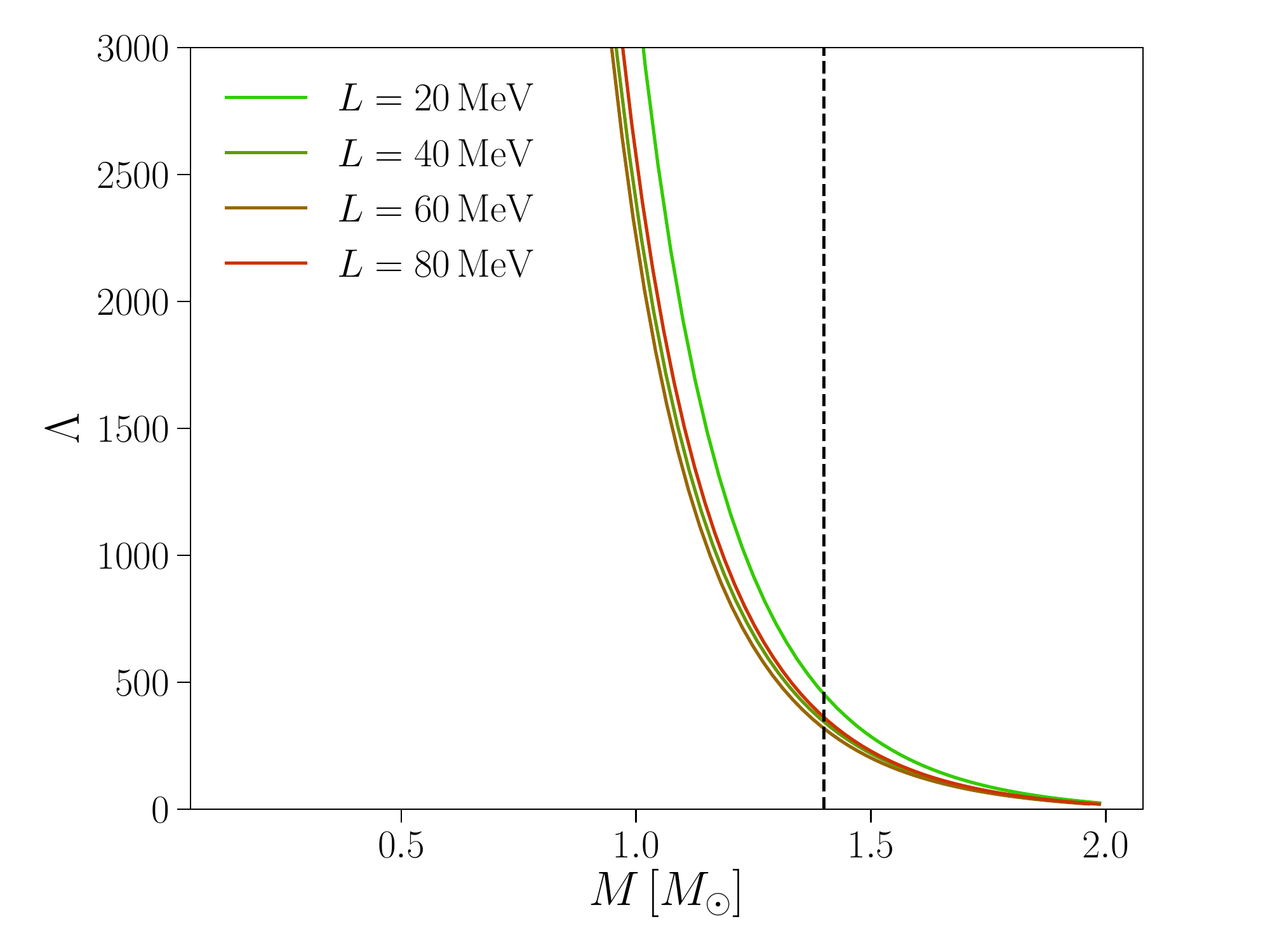}
\includegraphics[width=20pc]{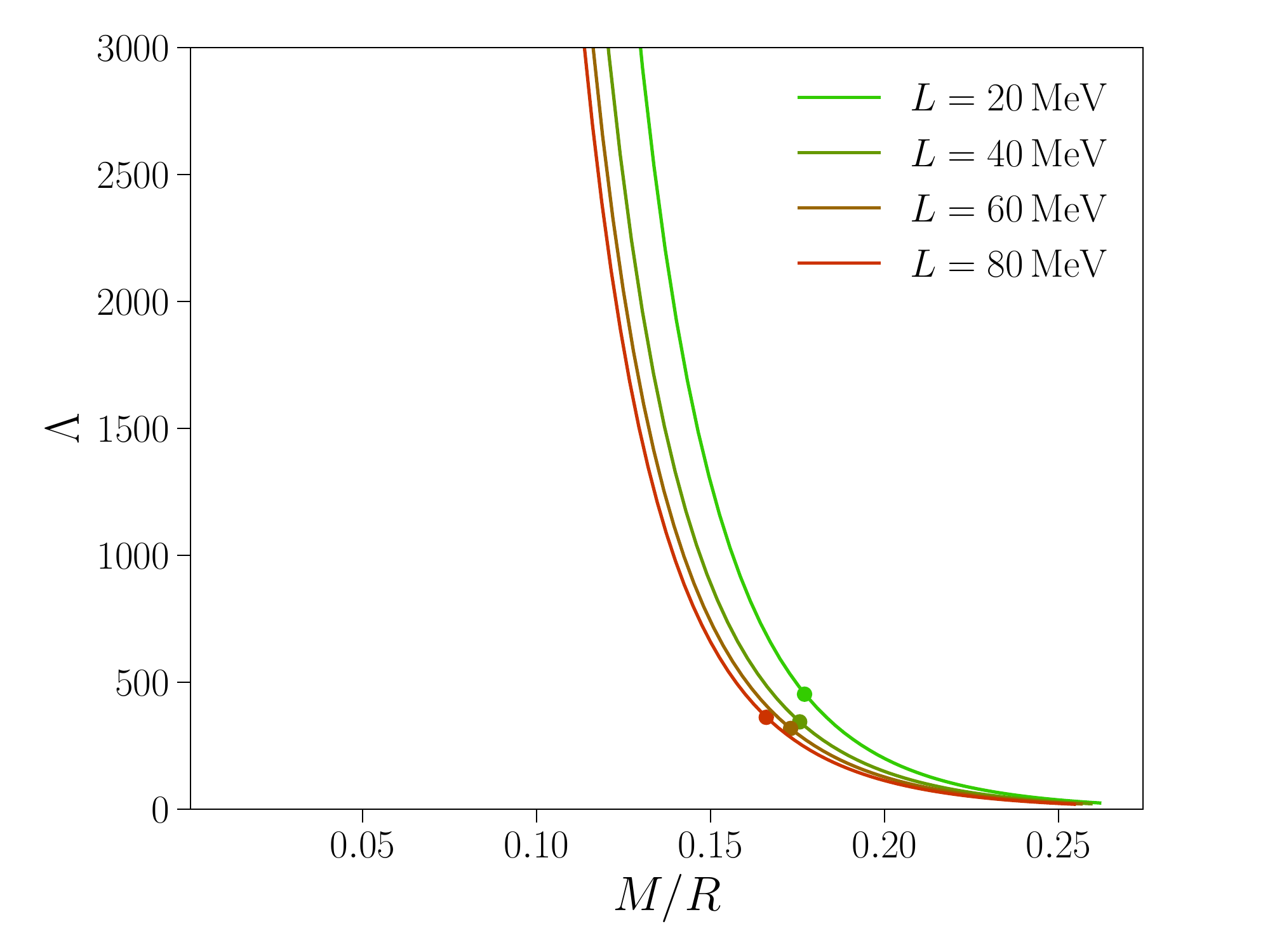}
\caption{Same with Figure 3, but for the tidal deformabilities.}
\end{figure}

The resulting mass-radius relations with $L$ = 20, 40, 60, 80 MeV are presented in Figure 2. They all fulfill the recent observational constraints of the two massive pulsars of which the masses are precisely measured~\citep{2013Sci...340..448A,2010Natur.467.1081D,2016ApJ...832..167F}. Since these four EOSs have the same incompressibility ($K = 240$ MeV) and symmetry energy ($E_{\rm sym} = 31$ MeV) but rather different symmetry energy slope $L$, it is clearly demonstrated that the radius sensitively depends on the symmetry energy slope with the maximum mass only slightly modified. It is the well accepted $R$ vs $L$ dependence mentioned in the introduction~\citep[e.g.,][]{2004Sci...304..536L,2001ApJ...550..426L,2006PhLB..642..436L}. A smaller $L$ (softer symmetry energy) leads to a smaller radius. The combined results for $\Lambda(1.4)$ are as shown in Figure 2. They all fulfill the GW170817 constraint of $\Lambda(1.4) \le 800$~\citep{2017PhRvL.119p1101A}. Different crust prescriptions could have influence on the resulting radius, we also test the crust dependence of tidal deformability by connecting the $L = 40$\,MeV EOS model with four other crust models collected in our previous work \citep{2016arXiv161008770L}. The resulting $\Lambda(1.4)$ is in the range of $324 \sim 344$, namely in the present study the crust influence of tidal deformability is limited around $6~\%$ for a $1.4\,M_\odot$ star.

We proceed to present in Figure 3 (Figure 4) the resulting Love numbers (tidal deformabilities) as a function of the mass and the compactness. The behaviours of $k_2$ and $\Lambda$ follow the above analysis and are similar with previous calculations~\citep[e.g.,][]{2010PhRvD..81l3016H,2010PhRvD..82b4016P}. In Figure 3, $k_2$ first increases then decreases with the mass and the compactness. In Figure 4, $\Lambda$ monotonously decreases with the mass and the compactness. The increase of $k_2$ and large values of $\Lambda$ for small masses (below $\sim 1.0 \,M_{\odot}$) are due to large radii and large portion of soft crust matter. If no crust is considered (e.g. an EOS described by a pure polytropic function), $k_2$ will decrease monotonously with mass and compactness as well~\citep{2010PhRvD..81l3016H}.

It's worthy noting that according to Figure 3, $k_2$ monotonically depends on $L$ for all the mass range, i.e., for a star with certain amount of mass (compactness), a larger $L$ leads to a smaller $k_2$. However, as already seen in Figure 2 from $\Lambda(1.4)$, this monotonic dependence doesn't hold for $\Lambda$, since $\Lambda$ is normalized with a factor of $R^5$. Hence the differences in radius (according to the $R$ vs $L$ relation mentioned before) will scatter the dependence of $\Lambda$ on $L$.

\begin{figure}\label{lambdal}
\includegraphics[width=40pc]{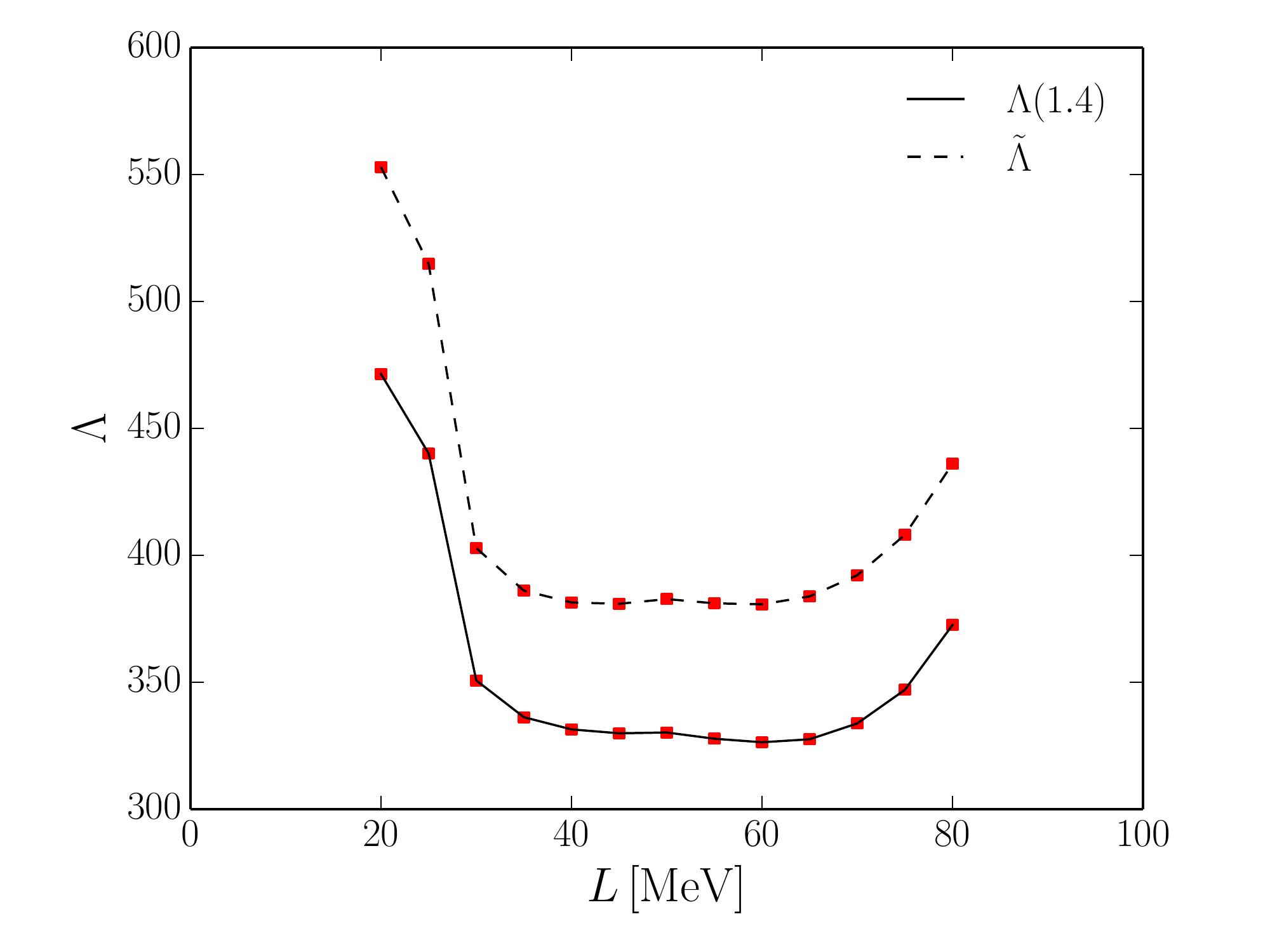}
\caption{$L$ dependence of tidal deformability: (In solid) the tidal deformability for a $1.4\,M_\odot$ star and (in dashed) the mass-weighed tidal deformability $\tilde{\Lambda}$ of a binary system with a chirp mass of 1.188$\,M_{\odot}$, and mass ratio of 0.7~\citep{2017PhRvL.119p1101A}.}
\end{figure}

To understand better the relation between $L$ and $\Lambda$, we present in Figure 5, for more $L$ values, the results of both $\Lambda(1.4)$ and the measured mass-weighed tidal deformability ($\tilde{\Lambda}$). A chirp mass of 1.188$\,M_\odot$ and mass ratio of 0.7\footnote{As pointed out by \citep[e.g.,][]{2018ApJ...852L..29R}, the mass-weighed tidal deformability is expected to be very weakly dependent on the mass ratio. Hence considering one mass ratio case should be representative enough for analysis.} are employed for the calculation of $\tilde{\Lambda}$ in a binary system. We can see that neither $\Lambda(1.4)$ nor $\tilde{\Lambda}$ shows correlation with $L$. In general, one expects large $\Lambda$ at large $L$ since a large star (i.e., large $R$) tends to be easily deformed. However, $\Lambda$ is anomalously large at small $L$ (i.e. $L\le30\,$MeV), which may be understood from the different smoothness behaviour in these cases at the crust-core matching interface, when core EOSs with different $L$ are matched to the same inner crust EOS of~\citet{nv}. The importance of matching interface has been pointed out also in our previous work \citep{2016arXiv161008770L} for the mass-equatorial radius relations of fast spinning NSs and deserve a better study in the future. The violation of monotonic dependence of $\Lambda$ on $L$ is particularly interesting in terms of observations. As a consequence, a measurement of $\tilde{\Lambda}$ by GW observations doesn't necessarily translate into measurement of radius of the NS, given the $R$ vs $L$ relation. Similar conclusion has been recognized also in the extended Skyrme-Hartree-Fock model~\citep{chen}. 

Finally in this section, we provide the tabulated EOS in Table 3 for our presently best model (the case of $L = 40$ MeV), with satisfying descriptions of vacuum nucleon properties ($r_N, m_N$), nuclear matter properties ($\rho_0$, $E/A$, $K$, $E_{\rm sym}$, $L$, $M_N^\ast$), and astrophysical observations ($M_{\rm TOV}, \Lambda(1.4)$). It is named as ``QMF18'' EOS. Without interpolation, the EOS data in Table 3 give $2.0805 M_{\odot}$ for $M_{\rm TOV}$, within an error of magnitude of $\sim 10^{-4}$: the complete data give $2.0809 M_{\odot}$.
 
\begin{table}\label{tab3}
\begin{center}
\caption{NS EOS for the QMF18 model newly introduced in this work.}
\begin{tabular}{c|c|c}\hline
 $\epsilon$~[g~cm$^{-3}$] & $P$~[erg~cm$^{-3}$] & $\rho_N$~[fm$^{-3}$]  \\ \hline
 0.13855E+15 & 0.79586E+33 & 0.082    \\
 0.14365E+15 & 0.85234E+33 & 0.085    \\
 0.15216E+15 & 0.95144E+33 & 0.090    \\
 0.16920E+15 & 0.11706E+34 & 0.100    \\
 0.18626E+15 & 0.14226E+34 & 0.110    \\
 0.20336E+15 & 0.17145E+34 & 0.120    \\
 0.22047E+15 & 0.20433E+34 & 0.130    \\
 0.27203E+15 & 0.33950E+34 & 0.160    \\
 0.32393E+15 & 0.55426E+34 & 0.190    \\
 0.37631E+15 & 0.87679E+34 & 0.220    \\
 0.42926E+15 & 0.13315E+35 & 0.250    \\
 0.48293E+15 & 0.19385E+35 & 0.280    \\
 0.53741E+15 & 0.27149E+35 & 0.310    \\
 0.59282E+15 & 0.36752E+35 & 0.340    \\
 0.64927E+15 & 0.48329E+35 & 0.370    \\
 0.70686E+15 & 0.62008E+35 & 0.400    \\
 0.76568E+15 & 0.77912E+35 & 0.430    \\
 0.82583E+15 & 0.96151E+35 & 0.460    \\
 0.88738E+15 & 0.11682E+36 & 0.490    \\
 0.95043E+15 & 0.13999E+36 & 0.520    \\
 0.10150E+16 & 0.16569E+36 & 0.550    \\
 0.10813E+16 & 0.19389E+36 & 0.580    \\
 0.11492E+16 & 0.22449E+36 & 0.610    \\
 0.12189E+16 & 0.25733E+36 & 0.640    \\
 0.12904E+16 & 0.29223E+36 & 0.670    \\
 0.13636E+16 & 0.32903E+36 & 0.700    \\
 0.14896E+16 & 0.39423E+36 & 0.750    \\
 0.16207E+16 & 0.46399E+36 & 0.800    \\
 0.17568E+16 & 0.53809E+36 & 0.850    \\
 0.18978E+16 & 0.61645E+36 & 0.900    \\
 0.20438E+16 & 0.69900E+36 & 0.950    \\
 0.21948E+16 & 0.78573E+36 & 1.000    \\
 0.25116E+16 & 0.97160E+36 & 1.100    \\
 0.28480E+16 & 0.11739E+37 & 1.200    \\
 0.32039E+16 & 0.13926E+37 & 1.300    \\ \hline
\end{tabular}
\end{center}
\end{table}

We also collect other new NS EOSs from various many-body techniques. These EOSs with their mass-radius relations are plotted in Figure~6. Their $L$ and $M_{\rm TOV}$ values are also shown in Table 4, with $\tilde{\Lambda}$ and various results for a $1.4\,M_\odot$ star: $R$(1.4), $M/R$(1.4) and $\Lambda$(1.4). The EOSs of NL3${\omega\rho}$, DDME2, and DD2 are from the RMF model~\citep{2016PhRvC..94c5804F}. The EOSs of DDRHF and DDRHF$\Delta$ is from the density-dependent relativistic Hartree-Fock (DDRHF) theory, with the later one extended to include $\Delta$-isobars~\citep{2016PhRvC..94d5803Z}. The Sly9 EOS is from the Skyrme functional~\citep{2016PhRvC..94c5804F}. The BCPM EOS, named after Barcelona-Catania-Paris-Madrid energy density functional~\citep{2015A&A...584A.103S}, is based on the microscopic Brueckner-Hartree-Fock theory~\citep{1999nmne.conf....1B}. 

We see that the tidal deformability has a roughly positive relation with the symmetry energy slope $L$: A smaller $L$ usually leads to a smaller $\Lambda$(1.4). Nevertheless, $\Lambda$(1.4) depends not only on the saturation properties (like $L$), but also on the high-density part of EOS (imprinted on $M_{\rm TOV}$). This is clearly seen in the comparison of the NL3${\omega\rho}$ and DD2 cases, where with the same $L \sim 55$ MeV, $\Lambda$(1.4) drops from $925$ for NL3${\omega\rho}$ to $674$ for DD2, resulting from a much lowered $M_{\rm TOV}$ value in the DD2 case: $2.75$ vs $2.42 \,M_\odot$. We notice that, besides NL3${\omega\rho}$, the DDRHF results with a representative parameter set PKO3 are not consistent with the $\Lambda(1.4)\le800$ constraint of GW170817 for the low-spin prior, but are allowed by a more loosely constrained upper limit of 1400 for the high-spin prior~\citep{2017PhRvL.119p1101A}. Also, possible strange phase transitions (e.g., $\Delta$-isobars~\citep{2016PhRvC..94d5803Z}), soften the high-density EOS and lower the maximum static gravitational mass $M_{\rm TOV}$, leading to relatively small values of $\Lambda(1.4)$: 865 (for DDRHF) vs 828 (for DDRHF$\Delta$).

\begin{figure*}
\epsscale{1.0}
\plotone{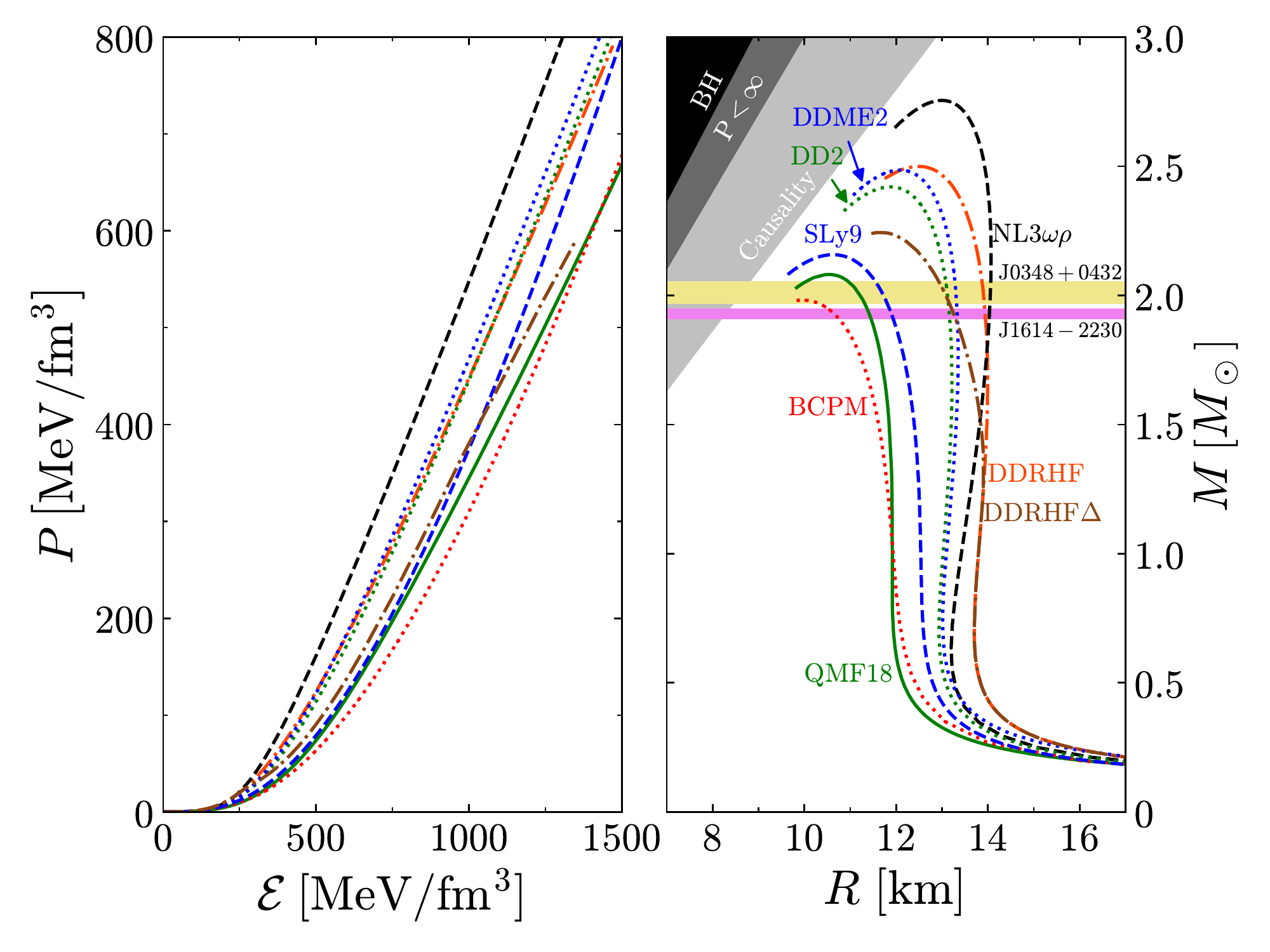}
\caption{Present new QMF18 EOS (left panel) and its mass-relation (right panel), to be compared with results of other recent NS EOSs from various many-body techniques: DDRHF, DDRHF$\Delta$~\citep{2016PhRvC..94d5803Z}, NL3${\omega\rho}$, DDME2, DD2, Sly9~\citep{2016PhRvC..94c5804F}, BCPM~\citep{2015A&A...584A.103S}. The mass measurements for two recent massive stars: PSR J1614-2230~\citep{2010Natur.467.1081D,2016ApJ...832..167F} and PSR J0348+0432~\citep{2013Sci...340..448A} are also shown. The shaded regions show the black hole limit, the Buchdahl limit and the causality limit, respectively.}\label{fig5}
\end{figure*}

\begin{table}\label{tab4}
\begin{center}
\caption{Radius, compactness and tidal deformability for a $1.4\,M_\odot$ star are provided for various advanced NS EOSs, together with their maximum static gravitational mass $M_{\rm TOV}$ and the symmetry energy slope $L$. In the last line we have also shown the range of $\tilde{\Lambda}$ for a binary system with chirp mass equal to 1.188$\,M_\odot$ and mass ratio in the range of $(0.7-1)$, which corresponds to the low spin case for GW170817. This calculation shows the consistency between the constraint in $\tilde{\Lambda}$ and $\Lambda(1.4)$. Further more, for NL3$\omega\rho$ EOS which possesses the largest value of $\Lambda(1.4)$ among all the EOSs, $\tilde{\Lambda}$ can actually be as small as 712 if the mass ratio of the system is 0.4 (which corresponds to the $90~\%$ credible range of the mass ratio in the high spin assumption case for GW170817), hence very close to the $90~\%$ credible upper limit for $\tilde{\Lambda}$ in the high spin case. Therefore, the possibility of this EOS wouldn't be clearly excluded if the high spin case in taken into account, as also seen in \citet{2018ApJ...857...12N}.}
\begin{tabular}{c|c|cc|ccccc}\hline
& QMF18 & DDRHF & DDRHF$\Delta$ & NL3${\omega\rho}$ & DDME2 & DD2 & Sly9 & BCPM \\  \hline
$M_{\rm TOV}~[M_\odot]$  & 2.08 & 2.50 & 2.24 & 2.75 & 2.48 & 2.42 & 2.16 & 1.98 \\
$L$ [MeV] & 40 & 82.99 & 82.99 & 55.5 & 51.2 & 55.0 & 54.9 & 52.96\\ \hline
$R$(1.4) [km] & 11.77 & 13.74 & 13.67 & 13.75 & 13.21 & 13.16 & 12.46 & 11.72 \\ 
$M/R$(1.4) & 0.1756 & 0.1505  & 0.1512 & 0.1503 & 0.1566 & 0.1571 & 0.1660 & 0.1765 \\
$\Lambda$(1.4) & 331 & 865 & 828 & 925 & 681 & 674 & 446 & 294\\  
$\tilde{\Lambda}$ & 381.4 - 388.4 & 948.7 - 993.4 & 900.8 - 962.9 & 1002.9 - 1056.3 & 747.8 - 782.7  & 747.9 - 777.3 & 
519.6 - 524.3 & 353.9 - 1056.3 \\ \hline
\end{tabular}
\end{center}
\end{table}

\section{Summary}

In the era of gravitational wave astronomy, the unknown EOS of supranuclear matter could soon be understood thanks to accumulating studies on gravity, astrophysics and nuclear
physics. The present work timely constructs a new EOS for NSs in the quark level, respecting all available constrains from terrestrial nuclear laboratory experiments and astrophysical observations, including the recent GW170817 constraint on the tidal deformability.

We employ the QMF model, where constituent quarks are confined by a harmonic oscillator confining potential. We first determine the quark potential parameter by reproducing properties of the nucleon in free space. Corrections due to center-of-mass motion, quark-pion coupling, and one gluon exchange are included to obtain the nucleon mass. Then the many-body nucleonic system is studied in the mean-field level, with the meson coupling constants newly fitted by reproducing the empirical saturation properties of nuclear matter, including the recent determinations of symmetry energy parameters. The predicted star properties can fulfill the recent two-solar-mass constraint and the 800 constraint for the dimensionless tidal deformability of a 1.4 $M_{\odot}$ star.

In particular, we explore the relation of the tidal deformability with an uncertain parameter of the symmetry energy slope at saturation. The discussions are done not only for modifying the slope value in its empirical range in one model, but also for comparing results of various many-body techniques. We find no evidence for a simple relation between the symmetry energy slope (hence the radius) and tidal deformability (either $\Lambda(1.4)$ or $\tilde{\Lambda}$). Consequently, claims regarding constraining NS radius with tidal deformablity measurements should be considered with caution. 

For future perspective, along this line, we can make detailed studies for tidal deformability on the interplay of the saturation parameters with various possible strangeness phase transition at higher densities (usually above $2\rho_0$), e.g., hyperons, kaon condensation, $\Delta$-isobars~\citep{2011PhRvC..83b5804B,2014PhRvC..89b5802H,2006PhRvC..74e5801L,2007ChPhy..16.1934L,2010PhRvC..81b5806L,2016PhRvC..94d5803Z}. We can also extend the present study to a unified treatment of both the hadron phase and the quark phase, for exploring better the quark deconfinement phase transition in dense matter and the properties of hybrid star~\citep[e.g.,][]{2015PhRvC..91c5803L}. An extend QMF18 EOS with unified crust and core properties will be useful as well for supernova simulation or pulsar studies. The pulsar properties can be predicted~\citep[e.g.,][]{2016ApJS..223...16L} and updated studies can be done for short gamma-ray bursts~\citep[e.g.,][]{2016PhRvD..94h3010L,2017ApJ...844...41L}. An extension of the current QMF model including a spherical bag for confinement is in progress~\citep{qmfbag}.

\begin{acknowledgements}
We would like to thank Bao-An Li and Antonios Tsokaros for valuable discussions; E. Z. is grateful for China Scholarship Council for supporting the joint Ph.D training project. The work was supported by the National Natural Science Foundation of China (No. U1431107).
\end{acknowledgements}

\end{document}